\documentclass[showpacs,prd,aps,12pt]{revtex4}

\usepackage{graphics}
\usepackage{psfig}
\usepackage{epsf}
\usepackage{epsfig}

\begin{document}

\def\Journal#1#2#3#4{{#1} {\bf #2}, #3 (#4)}

\def\NCA{\em Nuovo Cimento}
\def\NIM{\em Nucl. Instrum. Methods}
\def\NIMA{{\em Nucl. Instrum. Methods} A}
\def\NPB{{\em Nucl. Phys.} B}
\def\PLB{{\em Phys. Lett.}  B}
\def\PRL{\em Phys. Rev. Lett.}
\def\PRD{{\em Phys. Rev.} D}
\def\PRC{{\em Phys. Rev.} C}
\def\ZPC{{\em Z. Phys.} C}

\def\st{\scriptstyle}
\def\sst{\scriptscriptstyle}
\def\mco{\multicolumn}
\def\epp{\epsilon^{\prime}}
\def\vep{\varepsilon}
\def\ra{\rightarrow}
\def\ppg{\pi^+\pi^-\gamma}
\def\vp{{\bf p}}
\def\ko{K^0}
\def\kb{\bar{K^0}}
\def\al{\alpha}
\def\ab{\bar{\alpha}}
\def\be{\begin{equation}}
\def\ee{\end{equation}}
\def\bea{\begin{eqnarray}}
\def\eea{\end{eqnarray}}
\def\CPbar{\hbox{{\rm CP}\hskip-1.80em{/}}}

\def\lsim{\mathrel{\rlap{\lower4pt\hbox{\hskip1pt$\sim$}}
    \raise1pt\hbox{$<$}}}         
\def\gsim{\mathrel{\rlap{\lower4pt\hbox{\hskip1pt$\sim$}}
    \raise1pt\hbox{$>$}}}  

\newcommand\noi{\noindent} 
\newcommand\la{\langle}
\newcommand\eps\varepsilon

\hfill LA-UR-03-1443

\title{Investigating the Drell-Yan transverse momentum distribution
in the color 
dipole approach} 
\author{ M. A. Betemps
$^{1,a}$\footnotetext{$^1$E-mail:mandrebe@if.ufrgs.br},
 M. B. Gay Ducati
$^{2,a}$\footnotetext{$^2$E-mail:gay@if.ufrgs.br},  
M. V. T. Machado 
$^{3,a,b}$\footnotetext{$^3$E-mail:magnus@if.ufrgs.br}  
and J.\ Raufeisen $^{4,c}$\footnotetext{$^4$E-mail:jorgr@lanl.gov} }

\affiliation{ $^a$ Instituto de F\'{\i}sica, Universidade
Federal do Rio Grande do Sul\\ Caixa Postal 15051, CEP 91501-970, Porto
Alegre, RS, BRAZIL\\
$^b$ Instituto de F\'{\i}sica e Matem\'atica, Universidade Federal de
Pelotas\\
Caixa Postal 354, CEP 96010-090, Pelotas, RS, BRAZIL\\
$^c$ Los Alamos National Laboratory, MS H846, Los Alamos,
New Mexico 87545}

\vspace*{1cm}

\begin{abstract}
\vspace*{0.5cm}
\centerline{\bf Abstract}
\noindent
We study the influence of unitarity corrections on
the Drell-Yan transverse momentum distribution
within the color dipole approach. 
These unitarity corrections are implemented through the multiple
scattering Glauber-Mueller approach, which is contrasted with a
phenomenological saturation model. 
The process is analyzed for the center of mass
energies of the Relativistic
Heavy Ion Collider (RHIC, $\sqrt{s}=500$ GeV) and of the Large Hadron Collider
(LHC, $\sqrt{s}=14$ TeV). 
In addition, the results are extrapolated down to
current energies of proton-proton
collisions, where non-asymptotic corrections to the dipole approach
are needed.
It is also shown that
in the absence of saturation, the dipole approach can be 
related to the QCD Compton process.
\end{abstract}

\pacs{ 13.85.Qk; 12.38.Bx; 12.38.Aw. }
\maketitle

\clearpage

\section{Introduction}

\noi

The high energies available in the hadronic reactions at RHIC (BNL
Relativistic Heavy Ion Collider) and to be reached at LHC (CERN Large
Hadron Collider) will provide a better knowledge concerning parton
saturation. In such a kinematical region
the production of
massive lepton pairs in hadronic collisions (Drell-Yan (DY)
process \cite{originalDY}) can be used to investigate the high parton
density limit, since it is a clean reaction probing the gluon distribution
through the QCD Compton process. 
In particular, the Drell-Yan transverse momentum ($p_T$) distribution
can be expected to be sensitive to saturation effects.

\begin{figure}[b]
  \centerline{\scalebox{0.35}{\includegraphics{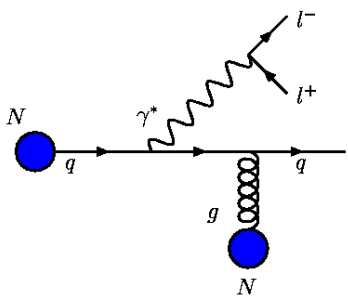}}}
    \center{
	{\caption{
      \label{dybrem}
      In the target rest frame,
      DY dilepton production looks like a bremsstrahlung. A quark 
      or an anti-quark from the
      projectile hadron scatters off the target color field 
	(denoted by a curly line)
	and radiates a
      photon ($\gamma^*$) with mass $M$ (before or after the quark scatters), 
which subsequently decays into the lepton pair 
($l^+l^-$). }  
    }  }
\end{figure}

Saturation and nuclear effects are most conveniently described within
the color dipole approach \cite{DYdipole2}, which is, in fact,
especially suitable for this purpose (see
Refs.~\cite{PRD66,NuclRauf,ptprc59,GelisJamal} for some applications).
The dipole approach is applicable only at high energies, and it is
formulated
in the target rest frame, where the DY process looks like a bremsstrahlung
of a virtual photon decaying into a lepton pair (see
Fig.~\ref{dybrem}). 
The advantage of this formalism is that
the DY cross section can be written in terms of the same
color dipole cross section as small-$x$ Deep Inelastic
Scattering (DIS).  
Although diagrammatically no dipole is present in bremsstrahlung, the dipole
cross section arises from the interference of the two bremsstrahlung 
diagrams, see Ref.~\cite{PRDrauf} for a detailed derivation. 
The cross section for a
radiation of a virtual photon from a quark scattering on a nucleon ($N$) can
be written in a factorized form as \cite{DYdipole2},
\begin{eqnarray}
\frac{d \, \sigma_{T,L}(qN\to \gamma^* X)}{d\ln\alpha} =\int \, d^2 r_{\perp}
\,  |\Psi^{T,L}_{\gamma^* q}(\alpha,r_{\perp})|^2  \,\sigma_{dip}
(\alpha r_{\perp}), 
\label{dycsalpha}
\end{eqnarray}
where $\sigma_{dip}$ is the same dipole cross section as in DIS, which
should take into account non-perturbative and saturation effects at
high energy \cite{PRD66}.  The energy
dependence of $\sigma_{dip}$ is not explicitly written out.  Here,
$r_{\perp}$ is the photon-quark 
transverse separation, and the argument of the dipole cross section,
$\alpha r_{\perp}$, is the displacement of the projectile quark in
impact parameter space due to the radiation of the virtual photon,
different to the DIS case, where the dipole separation is just
$r_{\perp}$. The $\Psi^{T,L}_{\gamma^* q}$ are the light-cone wave
functions for radiation of a transversely ($T$) or longitudinally
($L$) polarized photon (see e.g.\ Ref.~\cite{PRDrauf} for
explicit expressions).  While the light-cone wave functions are
calculable in perturbation theory, the dipole cross section can be
determined only with input from experimental data.

The goal of this work is to investigate the influence of unitarity
corrections on the DY dilepton $p_T$ distribution, describing these
unitarity corrections by the multiple scattering Glauber-Mueller
approach \cite{AGL} and including them into the dipole cross
section. The results are contrasted with the QCD improved
phenomenological saturation model of $\sigma_{dip}$,
Ref.~\cite{newGBW}, which quite successfully describes DIS and
diffractive DIS data.  In lines of a previous work \cite{PRD66}, here
it is investigated the role of the $\gamma^* q$ wave functions in the
$p_T$ distribution, characterizing the relation between dipole sizes
and transverse momentum.  A striking advantage of the color dipole
picture is a finite cross section for the lepton pair $p_T$
distribution at small $p_T\to 0$, even in the leading order
calculation, feature associated with the saturation encoded in
the dipole cross section.  In the conventional parton model, the
perturbative calculation of ${\cal O}(\alpha_s)$ yields a divergence
at $p_T=0$, and one has to resume large logarithms, $\ln\,(p_T^2/M^2)$,
in an appropriated scheme \cite{cososte}, in order to obtain a
physically sensible result.

The large amount of low Bjorken-$x$ DIS available data allows to
constrain the dipole cross section at very high energies and to
calculate the DY cross section without additional free
parameters. However, in the current energies of the hadronic colliders
there are non-asymptotic corrections to the dipole cross section,
which have to be taken into account, in order to describe experimental
data.  Therefore, one also introduces a parameterization for that
contribution, which is negligible already at RHIC
energies. In addition, we show how the dipole
approach for the DY $p_T$ distribution is related to the QCD Compton
process, which contributes at order $\alpha_s$ to the conventional
parton model of DY dilepton production. The two
approaches are equivalent in a certain limit.

\section{Relating dipole approach and parton model of high $p_T$
dilepton production}\label{sec:relate}

\begin{figure}[t]
  \centerline{\scalebox{0.8}{\includegraphics{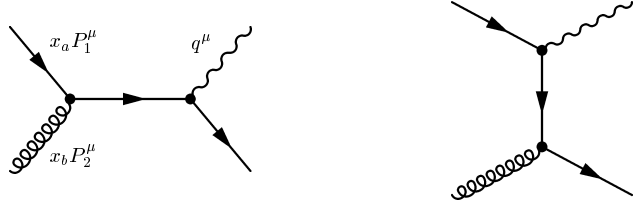}}}
    \center{
	{\caption{
      \label{fig:compton} Production of massive photons through the QCD
Compton process. The subsequent decay of the $\gamma^*$ (wavy line)
into the dilepton pair is not shown here. Curly lines denote gluons,
quarks are represented by lines with arrows.}  
    }  }
\end{figure}

Although the dipole approach and the next-to-leading order (NLO) 
parton model have been compared numerically in Ref.~\cite{PRDrauf},
one may still wonder, how these two approaches can be related to one
another analytically. This will be the topic of the present section.
Using the leading order expression \cite{fs},
\be
\sigma_{dip}(x,r_{\perp}) = \frac{\pi^{2} \,
\alpha_s}{3}\,\, r_{\perp}^{2}\,x\,G(x)\,,
\label{eq:fs} 
\ee for the dipole cross section, it can be demonstrated how the
dipole approach for high $p_T$ dilepton production is related to the
QCD Compton process, see Fig.~\ref{fig:compton}.  In
Eq.~(\ref{eq:fs}), $x\,G(x)$ is the density of gluons with momentum
fraction $x$ in a nucleon, and $\alpha_s$ is the strong coupling
constant.  First, we shortly review the formulas for high $p_T$
dilepton production in the dipole approach and in the parton model,
before we show how they can be translated into each other.

In the dipole approach, the DY transverse momentum distribution is given
by \cite{RaufThesis},
\bea\nonumber
\lefteqn{\frac{d^3\sigma(pp\to l^+l^-X)}{dydM^2dp_T^2}} \\
\nonumber
&=&\frac{\alpha_{em}}{3M^2}x_1
\int\limits_{x_1}^{\alpha_{max}}\frac{d\alpha}{\alpha^2}
\sum\limits_{q=1}^{N_f}e_q^2
\left[q\left(\frac{x_1}{\alpha}\right)
+\bar q\left(\frac{x_1}{\alpha}\right)\right]\\
\nonumber
&\times&\int d^2r_\perp d^2r^\prime_\perp 
e^{i\vec p_T\cdot\left(\vec r_\perp-\vec r^\prime_\perp \right)}
\left[
\Psi^T_{\gamma^*q}(\alpha,r_\perp)\Psi^{T*}_{\gamma^*q}(\alpha,r^\prime_\perp)
+\Psi^L_{\gamma^*q}(\alpha,r_\perp)\Psi^{L*}_{\gamma^*q}(\alpha,r^\prime_\perp)
\right]\\
\label{eq:dipolept}
&\times&\frac{1}{2} \left[\sigma_{dip}(x,\alpha
r_{\perp})+\sigma_{dip}(x,\alpha r^\prime_{\perp})
-\sigma_{dip}(x,\alpha \left|\vec r_{\perp}-\vec
r^\prime_{\perp}\right|)\right], \eea where the quark (antiquark)
distributions in the projectile are denoted by $q$ ($\bar q$
respectively). The usual definitions of the kinematic variables are
employed, i.e., 
\be x_1=\frac{2P_2\cdot q}{s}\quad,\quad
x_2=\frac{2P_1\cdot q}{s}, \ee where $q$ is the four momentum of the
virtual photon ($M^2=q^2$), and $P_{1,2}$ are the four momenta of the
projectile ($1$) and target ($2$) hadron.  By evaluating the scalar
product for $x_1$ in the target rest frame, it is easy to show, that
the projectile parton distributions in Eq.~(\ref{eq:dipolept}) are probed at
momentum fraction $x_1/\alpha$, where $\alpha$ is the 
momentum fraction taken by the photon from the projectile quark.
Furthermore, $p_T$ is the transverse momentum of the $\gamma^*$ in a
frame with the $z$-axis parallel to the projectile \ quark, and 
\be
y=\frac{1}{2}\ln\left(\frac{x_1}{x_2}\right) 
\ee 
is the rapidity of the photon. In addition, 
\be
\eta^2=\left(1-\alpha\right)M^2+\alpha^2m_q^2.  
\ee 
The quark mass
$m_q$ is set to 0 in this section.  The upper limit of the
$\alpha$-integration in Eq.~(\ref{eq:dipolept}) is determined from the
condition that the invariant mass of the final state cannot exceed
the total available center of mass (c.m.) energy of the projectile
quark-target nucleon system, i.e., \be
\frac{x_1s}{\alpha}\ge\frac{p_T^2+\eta^2}{\alpha\left(1-\alpha\right)}\quad
\ra\quad \alpha_{max}=1-\frac{p_T^2}{x_1s-M^2}, \ee where $\sqrt{s}$
is the hadronic c.m.\ energy.  In the high energy approximation,
$\alpha_{max}=1$ for $s\to\infty$.  In this section, however, we
work with the exact value of $\alpha_{max}$.  

With $\sigma_{dip}$ given by Eq.~(\ref{eq:fs}), the integrals over
$r_\perp$ and $r^\prime_\perp$ in Eq.~(\ref{eq:dipolept})
can be performed analytically with the result \cite{RaufThesis},
\bea\label{eq:dipoler2}\nonumber
\lefteqn{\left(\frac{d^3\sigma(pp\to l^+l^-X)}{dydM^2dp_T^2}\right)_{r^2_\perp-{\rm approx.}}}\\
\nonumber
&=&\frac{\alpha^2_{em}\alpha_s}{9M^2}x_1
\int\limits_{x_1}^{\alpha_{max}}d\alpha
\sum\limits_{q=1}^{N_f}e_q^2
\left[q\left(\frac{x_1}{\alpha}\right)
+\bar q\left(\frac{x_1}{\alpha}\right)\right]xG(x)\\
&\times&\left\{
\left[1+\left(1-\alpha\right)^2\right]
\frac{p_T^4+\eta^4}{\left(p_T^2+\eta^2\right)^4}
+4M^2\left(1-\alpha\right)^2\frac{p_T^2}{\left(p_T^2+\eta^2\right)^4}
\right\}.
\eea
In order to obtain Eq.~(\ref{eq:dipoler2}), one has to assume that $xG(x)$
does not depend on $r_\perp$ through scaling violations.
Note also, that the $r^2_\perp$-approximation, Eq.~(\ref{eq:fs}), is 
applicable only at large $p_T$.

In the parton model, on the other hand, the high $p_T$ distribution of
DY dileptons produced via the QCD Compton process, see
Fig.~\ref{fig:compton}, is given by
\bea\nonumber\label{eq:pm}
\lefteqn{\left(\frac{d^3\sigma(pp\to l^+l^-X)}{dydM^2dp_T^2}\right)_{{\rm Compton }}}\\
\nonumber
&=&\frac{\alpha^2_{em}\alpha_s}{9M^2}\int\limits_{x_a^{min}}^{1} dx_a
\frac{x_ax_b}{x_a-x_1}\sum\limits_{q=1}^{N_f}e_q^2
\left\{\left[q\left(x_a\right)+\bar q\left(x_a\right)\right]G(x_b)
+G(x_a)\left[q\left(x_b\right)+\bar q\left(x_b\right)\right]\right\}\\
&\times& 
\frac{1}{\hat s^2}\left[-2M^2\frac{\hat t}{\hat s\hat u}
-\frac{\hat s}{\hat u}-\frac{\hat u}{\hat s}\right]
\eea
(see e.g.\ Ref.~\cite{field} for details). In Eq.~(\ref{eq:pm}), 
$x_a$ and $x_b$ are the momentum fractions of the colliding partons, and
\be
x_a^{min}=\frac{x_1-M^2/s}{1-x_2}.
\ee
Note that
at finite $p_T$, $x_{a,b}\neq x_{1,2}$, where $x_{1,2}=
\sqrt{\frac{p_T^2+M^2}{s}}e^{(+,-)y}$. 
The partonic Mandelstam variables,
${\hat s}$, ${\hat t}$, ${\hat u}$, are defined in terms of $x_a$, $x_b$ and
the four-momenta of the colliding hadrons, see Fig.~\ref{fig:compton}. 
In order to compare 
Eqs.~(\ref{eq:dipoler2}) and (\ref{eq:pm}), one has to express the 
partonic Mandelstam variables in terms of $\alpha$ and $p_T^2$,
\bea\label{eq:s}
\hat s=\left(x_aP_1+x_bP_2\right)^2
&=&\frac{p_T^2+\eta^2}{\alpha\left(1-\alpha\right)},\\
\label{eq:t}\hat t=\left(q-x_bP_2\right)^2
&=&-\frac{p_T^2}{1-\alpha},\\
\label{eq:u}\hat u=\left(q-x_aP_1\right)^2
&=&-\frac{p_T^2+\eta^2}{\alpha}.
\eea
Furthermore,
\be
x_a=\frac{x_1}{\alpha}\quad,\quad 
x_b=\frac{p_T^2+\eta^2}{\left(1-\alpha\right)p_T^2+\eta^2}\,x_2.
\ee
Inserting the expressions for the partonic Mandelstam variables, 
Eqs.~(\ref{eq:s}), (\ref{eq:t}) 
and (\ref{eq:u}), into Eq.~(\ref{eq:pm}), one obtains
a result very similar to Eq.~(\ref{eq:dipoler2}), except for
the combinations of parton distributions,
\bea\label{eq:dipoler3}
\nonumber
\lefteqn{\left(\frac{d^3\sigma(pp\to l^+l^-X)}{dydM^2dp_T^2}\right)_{{\rm Compton }}}\\
\nonumber
&=&\frac{\alpha^2_{em}\alpha_s}{9M^2}x_1
\int\limits_{x_1}^{\alpha_{max}}d\alpha
\sum\limits_{q=1}^{N_f}e_q^2
\left\{\left[q\left(x_a\right)+\bar q\left(x_a\right)\right]x_bG(x_b)
+G(x_a)x_b\left[q\left(x_b\right)+\bar q\left(x_b\right)\right]\right\}
\\
&\times&\left\{
\left[1+\left(1-\alpha\right)^2\right]
\frac{p_T^4+\eta^4}{\left(p_T^2+\eta^2\right)^4}
+4M^2\left(1-\alpha\right)^2\frac{p_T^2}{\left(p_T^2+\eta^2\right)^4}
\right\}.
\eea

When saturation effects are neglected, the dipole approach
reproduces that part of the QCD Compton contribution to DY, in which
the quark comes from the projectile and the gluon from the target. 
Thus, the dipole approach is valid, when the first term in the 
convolution of parton distributions in Eq.~(\ref{eq:dipoler3}) dominates.
This is the case at large rapidity and at small $x_b$, both
conditions fulfilled. 
The range of validity of the dipole approach can of course
only be established a 
posteriori. This is similar to the problem of determining the lowest 
scale at which perturbative QCD still works. The dipole approach is 
phenomenologically successful for values of $x_2<0.1$, though most
parameterizations of the dipole cross section are fitted only to 
DIS data
with Bjorken-$x<0.01$. 

Regarding the rapidity ($y$)
range, in which the dipole formulation 
can be applied, some guidance on the minimal value of $y$ can
be obtained from the numerical comparison of the dipole approach and the
next-to-leading order (NLO) parton model in Ref.~\cite{PRDrauf}. 
At RHIC energy $\sqrt{s}=500$ GeV, virtually
no deviations between the dipole approach and the NLO parton model have been
found for $y>0.5$ \cite{PRDrauf}. This means that one can safely compare the 
dipole approach to future DY measurements from the two PHENIX muon
arms \cite{phenix}.

On the other hand, the dipole approach takes into account several
effects that will be important at high energies. A realistic
parameterization of the dipole cross section includes gluon saturation,
which is not contained in the standard parton model. Moreover,
$\sigma_{dip}$ contains information about the transverse momentum
distribution of the target gluons, thereby is more complete
than the gluon distribution in the collinear
factorization approach. Finally, with a realistic parameterization
of the dipole cross section at large separation $r_\perp$, one can
apply Eq.~(\ref{eq:dipolept}) also at low $p_T$, while the conventional
parton model $p_T$ distribution, Eq.~(\ref{eq:pm}), is applicable only at 
very high $p_T\gsim M$.  

\section{Features of the Drell-Yan cross section in the color dipole approach}

\noi

In this section, it is investigated which distances $r_\perp$ in impact
parameter space are important for the 
Drell-Yan $p_T$ distribution.
For this purpose, the behavior of the weight function
for $\sigma_{dip}$ as function of $\rho=\alpha r_\perp$ for different
values of $p_T$ is studied.

Three of the four Fourier integrals in Eq.~(\ref{eq:dipolept})
can be performed analytically with the result \cite{NuclRauf},
\begin{eqnarray}
\frac{d\, \sigma^{DY}}{dM^2\, dx_{F}\,d^{2}p_{T}}= 
\frac{\alpha^2_{\rm{em}}}{6\,\pi^3 M^2}\,
\frac{1}{(x_{1} + x_{2})}
\int_0^\infty d\rho
W(\rho,p_T)\sigma_{dip}(\rho),
\label{dyhadrocs}
\end{eqnarray}
where the weight function $W(\rho,p_T)$ is given by
\begin{eqnarray}\nonumber
W(\rho,p_T) & = & \int_{x_1}^{1}\frac{d\alpha}{\alpha^2}\,
\frac{x_1}{\alpha}\sum\limits_{q=1}^{N_f}e_q^2
\left[q\left(\frac{x_1}{\alpha},M^2\right)
+\bar q\left(\frac{x_1}{\alpha},M^2\right)\right]\\
\nonumber&\times &
\left\{ [m_{q}^{2}\alpha^{4}+2M^2(1-\alpha)^2]
\left[\frac{1}{p_{T}^{2}+\eta^2}T_1(\rho)-\frac{1}{4\eta}T_2(\rho)\right] \right. \\
& & \left. +\,\,[1+(1-\alpha)^2]\left[\frac{\eta p_T}{p_{T}^{2}+\eta^2}
T_3(\rho)-\frac{T_1(\rho)}{2}+\frac{\eta}{4}T_2(\rho)\right] \right\}\,,
\label{wfr}
\end{eqnarray}
and the functions $T_i$ read,
\begin{eqnarray}
T_1(\rho) & = & \rho J_0(p_T \rho/\alpha)K_0(\eta \rho/\alpha)/\alpha\,,\\
T_2(\rho) & = & \rho^2 J_0(p_T \rho/\alpha)K_1 (\eta \rho/\alpha)/\alpha^2\,,\\
T_3(\rho) & = & \rho J_1(p_T \rho/\alpha)K_1 (\eta \rho/\alpha)/\alpha\,.
\end{eqnarray}
The functions $J_0$ and $J_1$ are the first class Bessel functions of order 
$0$ and $1$, whereas  $K_{0}$ and $K_{1}$ are the second class modified Bessel 
functions of order $0$ and $1$ (MacDonald functions). 

It was shown in Ref.~\cite{PRD66} that for the ($p_T$ integrated) mass
distribution, the wave functions select the small $\rho$ region.
Large values of $\rho\gsim 2/M$ are exponentiated by the functions $K_{0,1}$.
It should be stressed here that large dipole sizes correspond
to the non-perturbative sector of the reactions, whereas small size
configurations give the perturbative piece. 

\begin{figure}[t]
\begin{center}
\epsfig{file=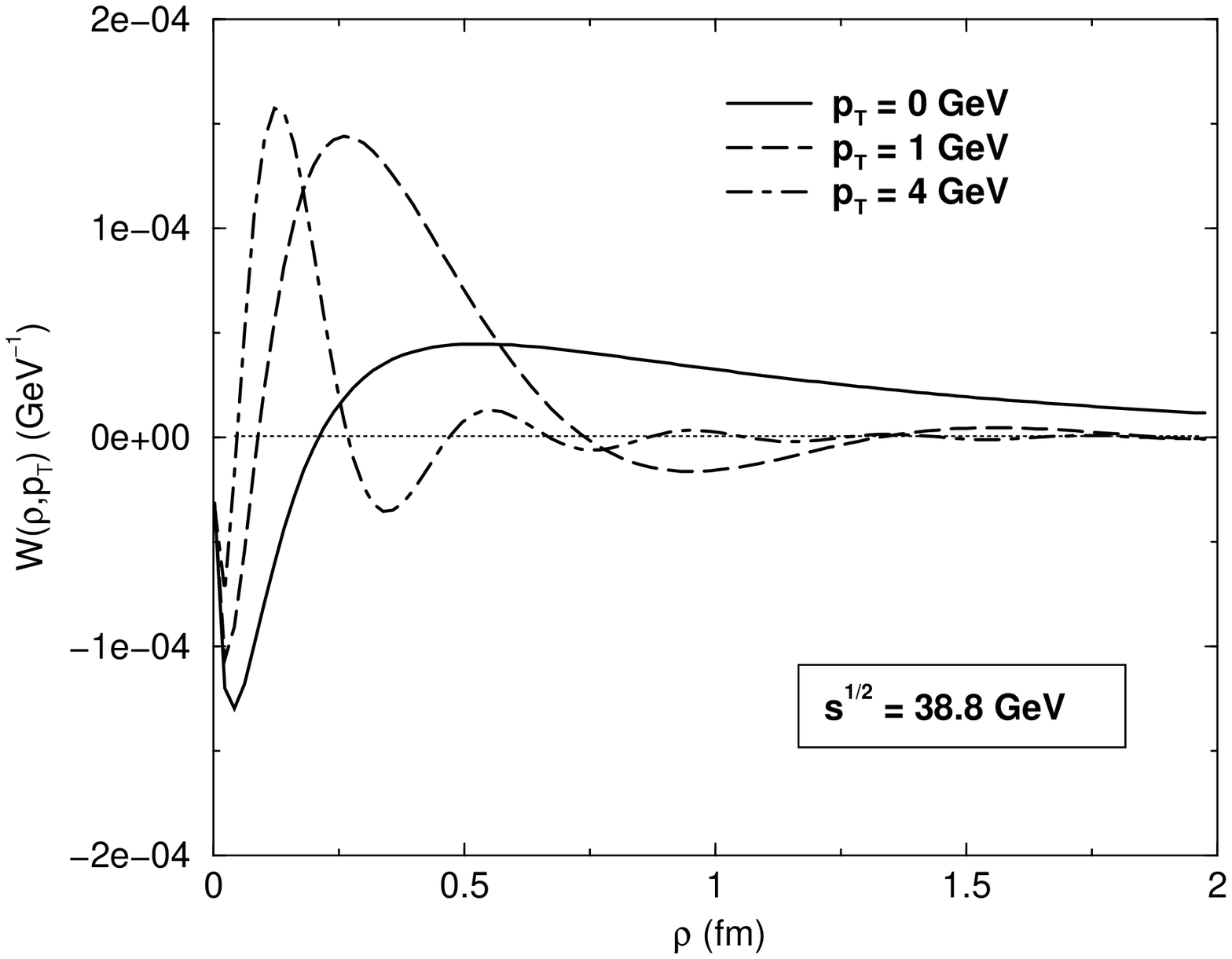,height=200.pt,width=200.pt}\hfill
\epsfig{file=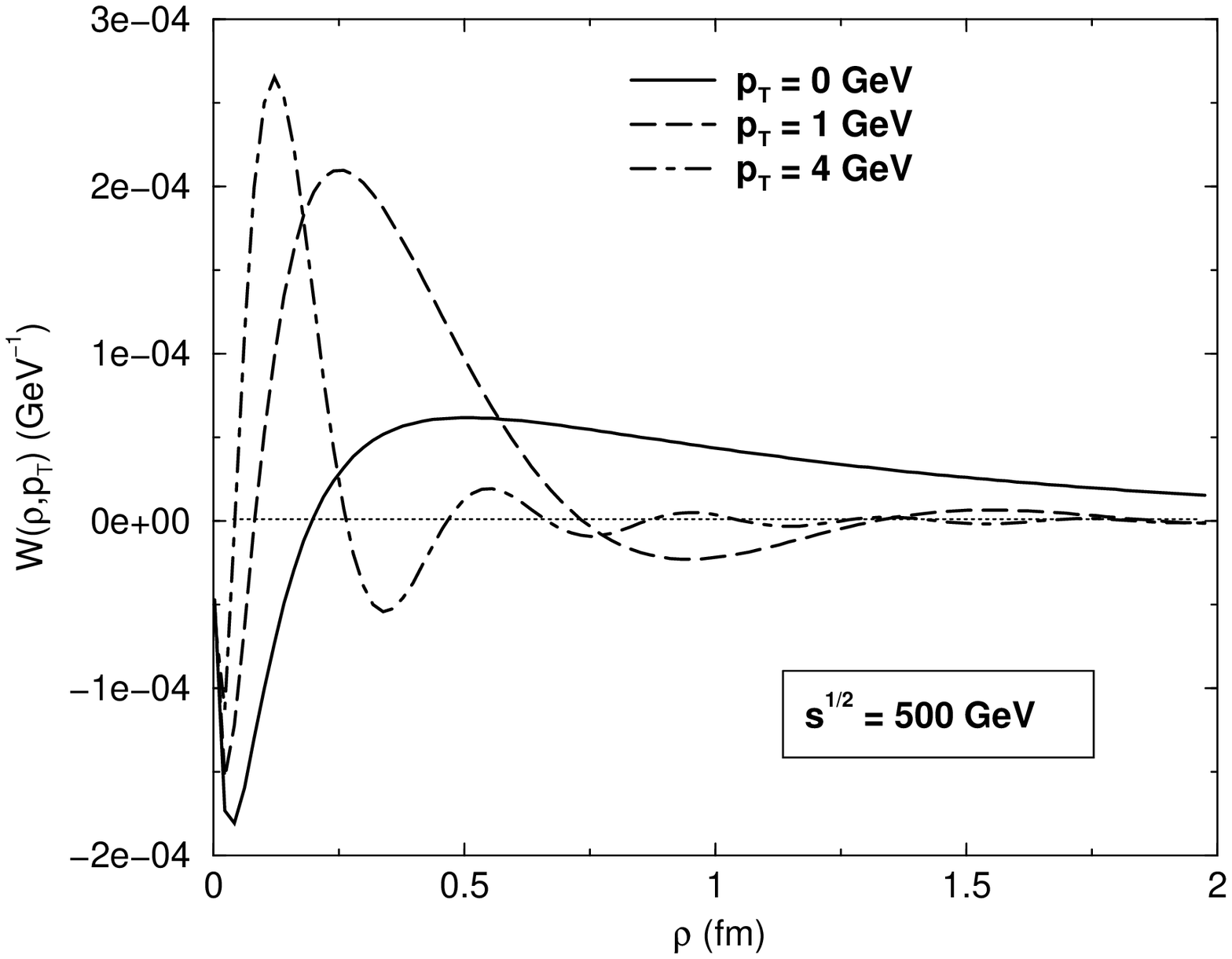,height=200.pt,width=200.pt}
\end{center}
\caption{The weight function $W(\rho,p_T)$ 
as a function of  $\rho$ for
different  $p_T$ at fixed  $x_F=0.625$ and $M=6.5$ GeV.}
\label{weightfunpt}
\end{figure}

In the
particular case of the dilepton $p_T$ distribution, a different
picture is designed.
In Fig.~\ref{weightfunpt}, we show
$W(\rho,p_T)$ as a function of the photon-quark transverse
separation $\rho$ for typical fixed lepton pair mass $M=6.5$ GeV
and Feynman-$x$ ($x_F=0.625$). The results are presented for two
center of mass energies: the plot on the left corresponds to
$\sqrt{s}=38.8$ GeV (available at the E772), whereas in the plot on
the right, $\sqrt{s}=500$ GeV (RHIC). For the effective light quark
masses, the value $m_q=0.2$ GeV was used.  Three different values
for the dilepton transverse momentum were selected, $p_T=0$, 1 and 4
GeV.

As can be seen from Eq.~(\ref{wfr}), the oscillating Bessel functions $J_i$
drive the behavior of the $W(\rho,p_T)$ as a function of
$\rho$. The following general picture can be drawn from the
plots: for large $p_T$ the large dipole size configurations get
suppressed, because $W(\rho,p_T)$ is rapidly oscillating. 
This suppression mechanism is different from the exponential suppression
of large dipole sizes in the case of the $p_T$ integrated cross section,
and complicates the numerical calculation of the $p_T$ distribution.
On the other hand, as $p_T$ decreases, large $\rho$
configurations become more important. 
The case $p_T=0$ is of particular interest, since the weight
function $W(\rho,p_T)$
selects very large dipole configurations and such a region is 
enhanced by increasing the energy. Therefore, the
non-perturbative sector of the process should drive the small $p_T$
regime.  On the other hand, the large $p_T$ behavior is almost
completely dominated by small dipole configurations \cite{Kop1}. These features
are exploited in the next section, where are also discussed the different
models that were employed for the dipole cross section.

\section{The dipole cross section}
\label{GMdipolesec}

\noi

The cross section for a small color dipole scattering on a 
nucleon can be
obtained from perturbative QCD \cite{fs}. 
However, there are large uncertainties stemming from
non-perturbative effects (infrared region) as well as  from
higher order and higher
twist corrections.  In the leading $\ln (1/x)$ approximation, the dipole
interacts with the target through the exchange of a
perturbative Balitsky-Fadin-Kuraev-Lipatov (BFKL) Pomeron,
described in terms of the ladder diagrams \cite{BFKL}. 
In the double logarithmic approximation, the BFKL equation \cite{BFKL} 
agrees with the evolution equation of Dokshitzer et al.\  \cite{DGLAP}
(hereafter DGLAP equation). 
In this limit, the dipole cross
section reads,
\begin{eqnarray}  
\sigma_{dip}(x,r_{\perp}) = \frac{\pi^{2} \,
\alpha_s}{3}\,\, r_{\perp}^{2}\,x\,G^{\mathrm{DGLAP}} (x,\tilde{Q}^2)\,,
\label{dglapd} 
\end{eqnarray} 
where $xG^{\mathrm{DGLAP}} (x,\tilde{Q}^2)$ is the usual DGLAP gluon
distribution at momentum fraction $x$ and virtuality scale
$\tilde{Q}^2=\lambda/r_{\perp}^{2}$. The factor $\lambda$ appearing in
the virtuality scale $\tilde{Q}^2=\lambda/r_{\perp}^{2}$, has been taken
as $\lambda=4$ \cite{PRD66}, although same magnitude values are equivalent at
leading logarithmic level \cite{McDermott}. The main feature of the
dipole cross section above is the color transparency property, i.e.,
$\sigma_{dip} \sim r_{\perp}^2$ as $r_{\perp} \rightarrow 0$. At large
dipole size, the dipole cross section should match the confinement
property $\sigma_{dip} \sim \sigma_0$. Concerning the large transverse
separation (non-perturbative sector), our procedure is to freeze the
$r^{2}_{\perp}$ in Eq.~(\ref{dglapd}) at a suitable scale larger than
$r^2_{\mathrm{cut}}$, which corresponds to the initial scale on the
gluon density perturbative evolution, $Q^2_0=4/r^2_{\mathrm{cut}}$.

At high energies, an additional requirement should be met: the growth
of the parton density (mostly gluons) has to be tamed, since an
uncontrolled increasing would violate the Froissart-Martin bound,
requiring the black disc limit of the target has to be reached at
quite small Bjorken $x$.  This feature can be implemented by using the
multiple scattering Glauber-Mueller approach (GM), which reduces the
growth of the gluon distribution by eikonalization in impact parameter
space \cite{AGL}.  Therefore, one substitutes $xG^{\mathrm{DGLAP}}$ in
Eq.~(\ref{dglapd}) by the corrected distribution including unitarity
effects, $x\,G^{GM}$.  A more extensive derivation of the GM dipole
cross section and the expression of $x\,G^{GM}$ can be found in the
Sec. III of the Ref.~\cite{PRD66}. Following previous work
\cite{PRD66}, one shall use $x=x_2$ as the energy scale in the dipole
cross section, since $x_2$ in DY is the analog of Bjorken-$x$ in
DIS. Note that $x=\alpha x_2$ was used in \cite{PRDrauf}, however, the
factor $\alpha$ has only a small numerical influence.

Once the dipole cross section is known, one can also calculate the DY
differential cross section, Eq.~(\ref{dyhadrocs}) integrated over
$p_T$, and compare it with the available data at small $x_2$. However,
the current data on DY reactions are measured in a kinematical region
where $x_2$ still takes rather large values, that is, $x_2
\simeq 0.1$ at $\sqrt{s}=38.8$ GeV, where the color dipole picture
reaches the limit of its validity. Therefore, in order to compare
the theory with these data, some procedure should be taken
to extend the applicability of the dipole cross section at large
$x_2$.  

Note, that the dipole cross section Eq.~(\ref{dglapd})
represents the asymptotic gluonic (Pomeron) contribution to the
process, and at large $x$ (low energy) a non-asymptotic quark-like
content should be included.  In the Regge theory language, this means
a Reggeon contribution, and therefore, we added 
the term
\cite{PRD66},
\begin{eqnarray}
\sigma^{I\!\!R}_{dip}=\sigma_{0}\, r_{\perp}^2 \,x^{\,0.425}\, (1-x)^3\,.
\label{reggpar}
\end{eqnarray} 
to the dipole cross section, Eq.~(\ref{dglapd}).

Using the expression above, good results are obtained in describing
the E772 data \cite{E772} on mass distribution with a Reggeon overall
normalization $\sigma_0 = 8$ \cite{PRD66}, reproducing similar results
considering the saturation model \cite{Kop1}. Nevertheless,
Eq.~(\ref{reggpar}) has a shortcoming when one calculates the dilepton
$p_T$ distribution: due to the fact that the weight function,
Eq.~(\ref{wfr}), selects large dipole configurations at small $p_T$
(see discussion in the previous section) the $\sim r_{\perp}^2$
behavior in the Reggeon dipole cross section produces a non-negligible
contribution at small $p_T$ even at RHIC energies.  Therefore,
Eq.~(\ref{reggpar}) was modified in order to cure this shortcoming and
preserve our previous results. The Reggeon contribution now reads,
\begin{eqnarray}
\sigma^{I\!\!R}_{dip}=\sigma_{0} \,r_{\perp}^2 \,x\,
q_{\mathrm{val}}\,(x,\tilde{Q}^2)\,,
\label{reggnew}
\end{eqnarray}
where the quantity $q_{\mathrm{val}}$ is the valence quark
distribution from the target and a reasonable description of the same E772
data is obtained with a value $\sigma_{0}=7$. The scaling violation
from the valence parton distribution takes care of the steep growing
on $r_{\perp}$, which is present in the simple parameterization of
Eq.~(\ref{reggpar}), removing the already mentioned shortcoming in the
$p_T$ distribution at high energies.  .

Our main goal here is to investigate the DY $p_T$ distribution, using
the GM dipole cross section.  However, for sake of comparison, this
analysis is contrasted with the phenomenological saturation
model of Bartels et al.\ (BGBK dipole cross section hereafter),
Ref.~\cite{newGBW}, which also includes the features of the
dipole cross section discussed above.  The model of Ref.~\cite{newGBW}
is a QCD improved version of the saturation model of
Ref.~\cite{GBW}. The new model explicitly includes QCD evolution, and
the dipole cross section is given by, 
\bea
\label{eq:newsat}
\sigma_{dip}(x,r_{\perp})= \sigma_{0}\left\{1-\exp\left(-\frac{\pi^2
r^2_{\perp} \alpha_{s}(\mu^2)xg(x,\mu^2)}{3\sigma_{0}}\right)\right\}\,, 
\eea
where the scale $\mu^2$ is assumed to have the form 
\bea
\mu^2=\frac{C}{r^2_{\perp}}+\mu_{0}^{2}.  
\eea 
The authors of
Ref.~\cite{newGBW} propose the following gluon distribution at initial
scale $Q_{0}^2=1$ GeV$^2$, 
\bea
xg(x,Q_{0}^2)=A_{g}x^{-\lambda_{g}}(1-x)^{5.6}.  
\eea 
Altogether,
there are five free parameters ($\sigma_{0}$, $C$, $\mu_{0}^2$,
$A_{g}$ and $\lambda_{g}$), which have been determined in
Ref.~\cite{newGBW} by fitting ZEUS, H1 and E665 data with $x<0.01$. In
this fit the parameter $\sigma_0$ is fixed at 23 mb during the fits as
in the original model, Ref.~\cite{GBW}.  Here, we employ fit 1 of
Ref.~\cite{newGBW}.

In Ref.~\cite{PRDrauf}, where 
the old saturation model of Ref.~\cite{GBW} was used,
the dipole approach was extrapolated to larger $x_2$ 
by introducing a threshold factor into the saturation scale, i.e.\
$Q^2_s \rightarrow Q^2_s\,(1-x_2)^5$. The factor $(1-x_2)^5$
is motivated from QCD counting rules and suppresses
the large $x_2$ contribution in the DY cross section. 
In our case, employing the GM or the 
BGBK dipole cross section, the large $x_2$
threshold factor is already included in the collinear gluon
distribution function. 

In addition, in Ref.~\cite{PRDrauf}, $(1-x_1)M^2$ was used as 
the virtuality scale, at which the projectile parton distribution
is probed, see Eq.~(\ref{dyhadrocs}). In this work, we shall use
$M^2$ instead. The factor $(1-x_1)$ is important only at large $x_1$,
but has no effect at midrapidity. In the next section we study
the dilepton transverse momentum distribution, making use of the
results obtained above for low energies

\section{The dilepton transverse momentum distribution}
\label{resdisc}

In this section, the DY dilepton transverse momentum distribution is
calculated, using the Glauber-Mueller dipole cross section,
Eq.~(\ref{dglapd}), and compared with the results obtained with the improved
saturation model, Eq.~(\ref{eq:newsat}).  We will consider
typical values for mass and $x_F$. The projectile structure
function employed was the LO GRV98 parametrization \cite{GRV98} to the
GM predictions and CTEQ5L \cite{CTEQ5} for the saturation model ones. 

Before doing that, some comments are in order. The unitarity effects
in the target will be significative at large rapidity 
$y=1/2\ln( x_{F}/x_{2}+1)$. In the central rapidity region
($y\simeq 0$) the effects in the projectile could be also
sizeable. In the last case, those effects in the quark distribution
are smaller than in the gluon content. Therefore they will be disregarded
in what follows.

\begin{figure}[b]
\begin{center}
\epsfig{file=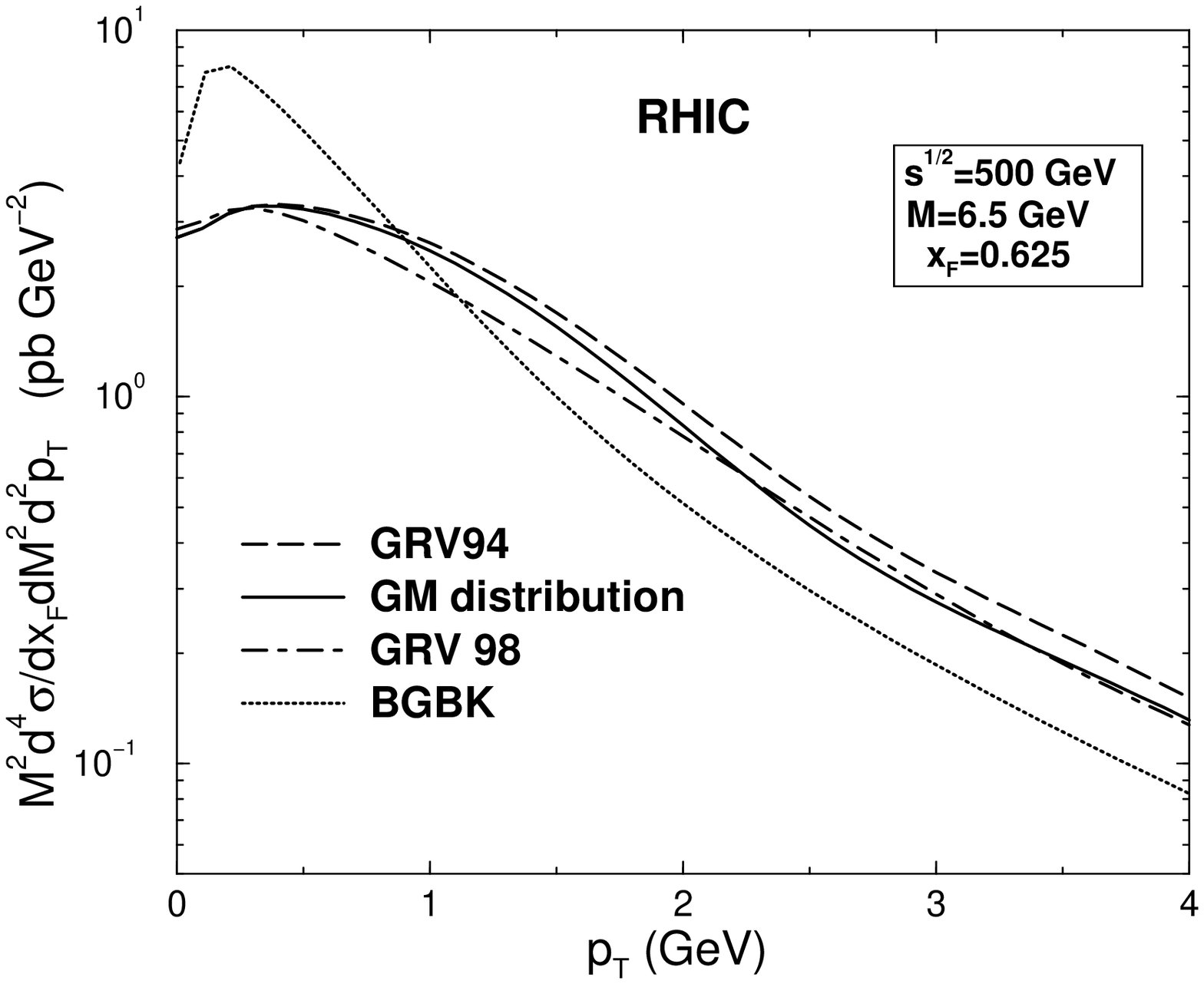,height=200.pt,width=220.pt}\hfill
\epsfig{file=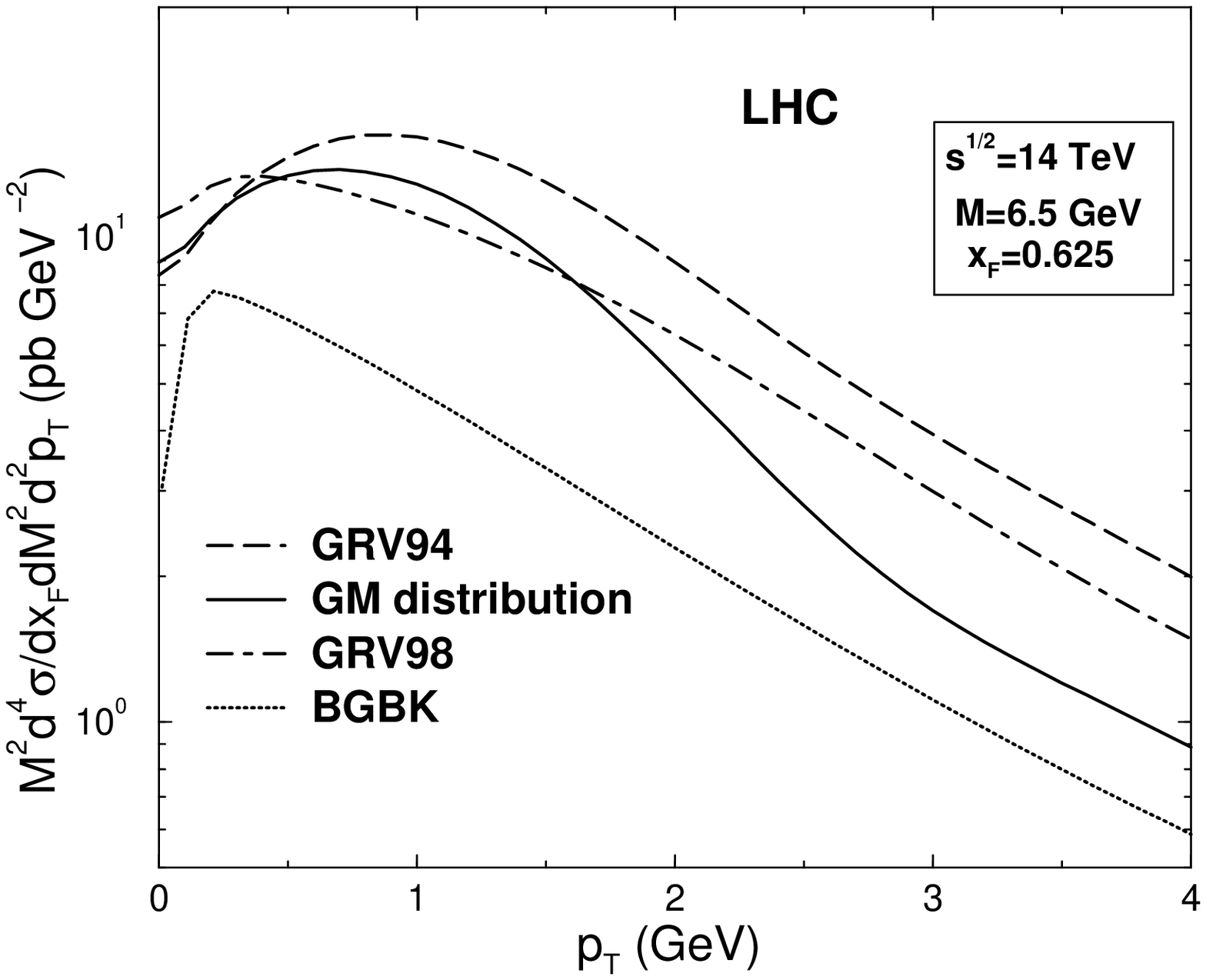,height=200.pt,width=220.pt}
\end{center}
\caption{The Drell-Yan dilepton transverse momentum $p_T$ distribution
at RHIC ($\sqrt{s}=500$ GeV) and LHC ($\sqrt{s}=14$ TeV). The solid
lines are the GM results including unitarity effects, the long-dashed
ones are the curves using GRV94 for the gluon distribution (without
unitarity effects) in the dipole cross section.  The dot-dashed curves
are the results obtained with the GRV98 gluon distribution (without
unitarity effects) in the dipole cross section and the dotted ones are
the results using the BGBK model.}
\label{fig4}
\end{figure}

In Fig.~\ref{fig4} results for the energies
from RHIC ($\sqrt{s}=500$ GeV) and LHC ($\sqrt{s}=14$ TeV) are shown
with $M=6.5$ GeV and $x_F=0.625$. 
At these energies and kinematics variables, the valence content is
completely negligible. 
We emphasize that the $x_F$ value considered above, is an extreme
  case, where the rapidity variable acquires large values for RHIC
  ($y\sim 3$) and LHC ($y\sim7$) energies.
In order to investigate the unitarity effects
for this observable, the following comparisons are performed:
The long-dashed curves are calculated with 
the dipole cross section, Eq.~(\ref{dglapd}),
without unitarity effects (denoted GRV94) using the GRV94 LO
parameterization \cite{GRV94} in calculating the dipole cross
section. The solid curves are the result including unitarity effects
with the same GRV94 parameterization as initial input. The use of this
parameterization is justified properly in Refs.~\cite{PRD66,mbgmvtm}.
The dot-dashed curves are calculated with 
the dipole cross section, Eq.~(\ref{dglapd}),
using as input the GRV98 parameterization for the gluon structure
function. The
aim of this comparison is to verify to what extent an updated
parameterization can absorb unitarity effects. 
It is verified that at RHIC energy, the unitarity
effects could be absorbed in the parameterization. However, at LHC
energy the situation is quite different, and the results are
completely distinct.  The deviation is important mostly at large
$p_T$, and as a general feature concerning the unitarity effects, those
corrections are significant at large transverse momenta and are
enhanced as the energy increases. 

As an additional comparison, we present curves from the improved
saturation model, Eq.~(\ref{eq:newsat}) (dotted lines in
Fig.~\ref{fig4}): at RHIC energies and quite small $p_T$, BGBK results
overestimates the GM one; however, at high $p_T$ the BGBK model
underestimates the GM predictions. At LHC energies, the BGBK
underestimates the GM results.  It is worth mentioning that until here the
analysis has been performed for fixed values of mass and $x_F$, which
implies that the values of the variable $x_2$ remain almost unchanged
in the analyzed $p_T$ interval.  The unitarity effects studied here
are calculated perturbatively, and thus they are more significant at
small $r$. At small $p_T$, large $r$ contributions are important and
even dominate in that region, which does not allow to observe the
saturation effects in a clear way. There, the confinement aspects of
the process are more important. In contrast, at large $p_T$ the main
contribution comes from the small $r$ region, which is sensible to the
inclusion of unitarity corrections to the process.

\begin{figure}[t]
\begin{center}
\epsfig{file=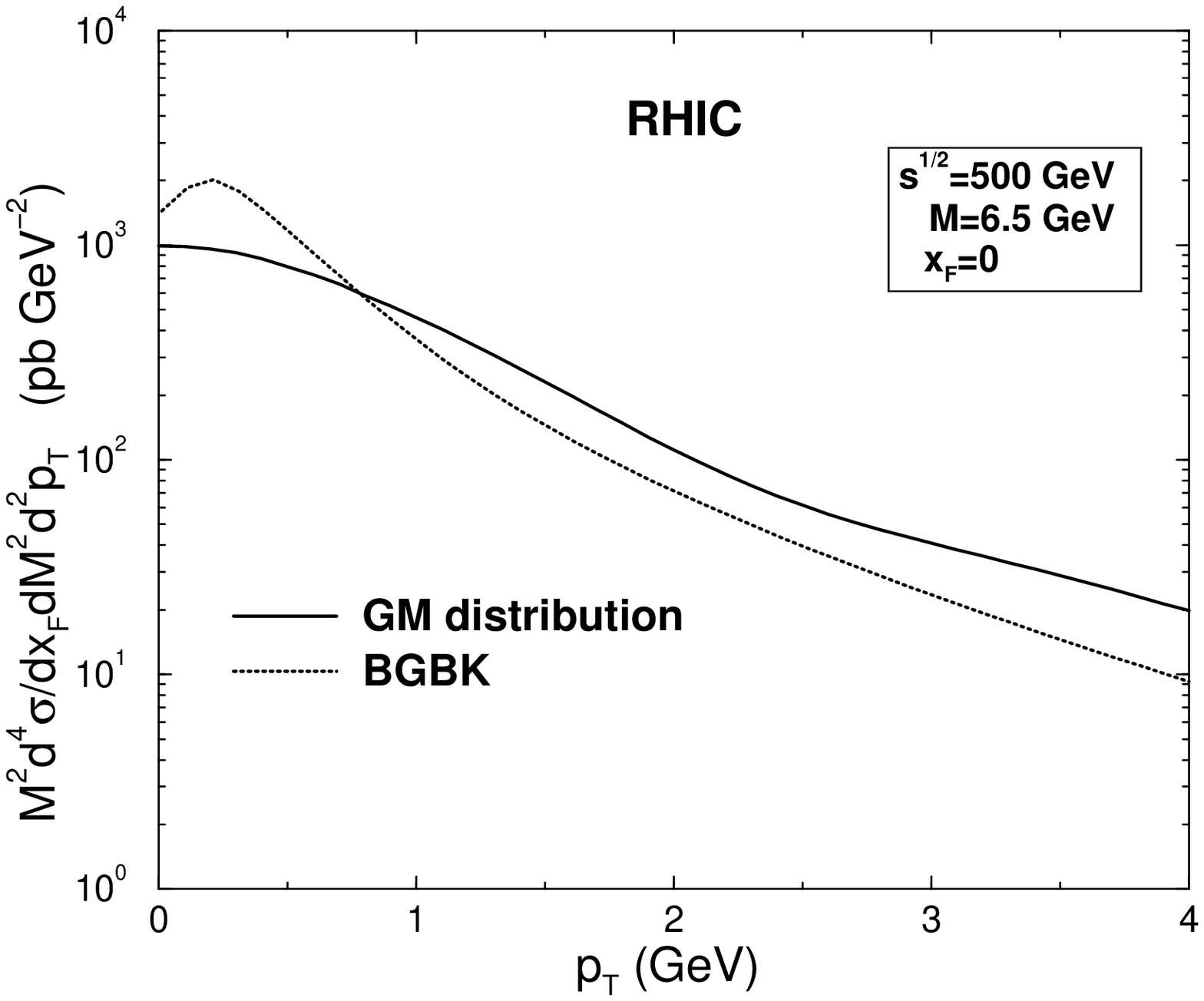,height=200.pt,width=220.pt}\hfill
\epsfig{file=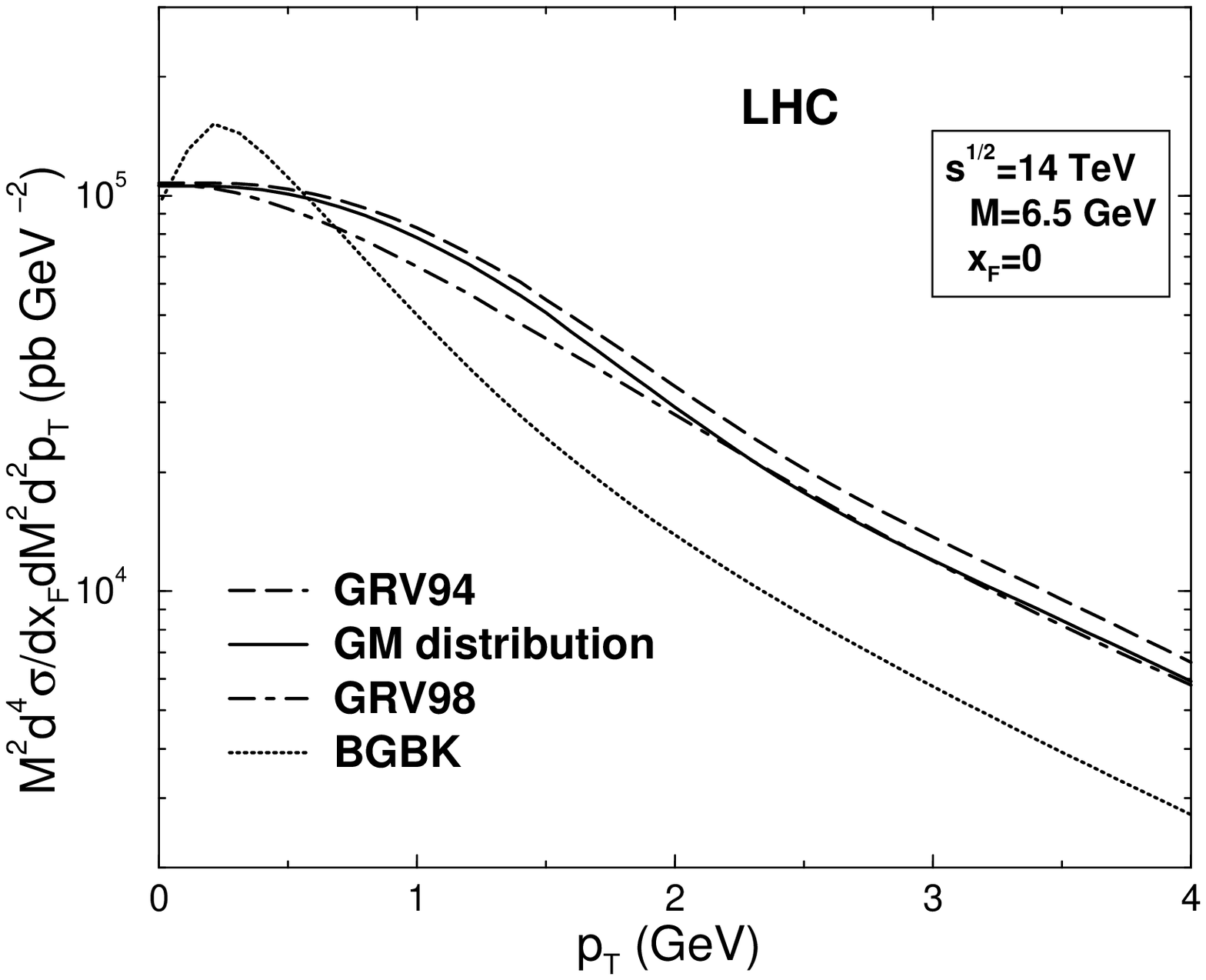,height=200.pt,width=220.pt}
\end{center}
\caption{The Drell-Yan dilepton transverse momentum $p_T$ distribution
at RHIC ($\sqrt{s}=500$ GeV) and LHC ($\sqrt{s}=14$ TeV). The
curves have the same meaning as in Fig.~\ref{fig4}. At RHIC
energy, only the GM and BGBK results are shown.}
\label{xf0}
\end{figure}

In order to perform estimates for more realistic values of the
kinematical variables, one considers that the DY measurements at these
colliders will be made predominantly in the central rapidity region
\cite{pajares},
i.e. at $x_F=0$, instead of a very forward direction. In
Fig.~\ref{xf0}, we present estimates for RHIC and LHC energies at
$x_F=0$. For RHIC, the results for the BGBK model (dotted line) and for
the GM approach (solid line) are shown, where the deviations are
larger at small $p_T$. For $x_F=0$, the deviations due to the
unitarity effects are smaller than for $x_F=0.625$, so only the GM
distribution is shown, since the results with the GRV94 and GRV98 are
almost the same as the one with the GM distribution. The results for LHC are
also presented. The unitarity effects are smaller
at $x_F=0$, because $x_2$ is larger at midrapidity than in the forward
direction.

\begin{figure}[t]
\begin{center}
\epsfig{file=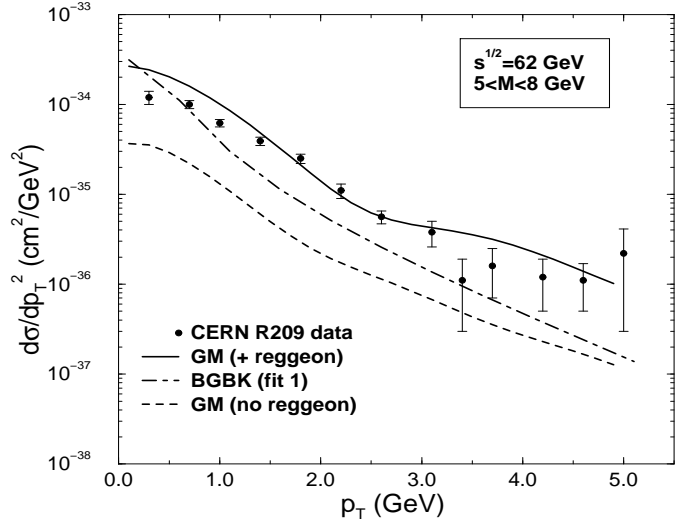,height=200.pt,width=250.pt}
\end{center}
\caption{The Drell-Yan differential cross section on $p_T$ at energy
$\sqrt{s}=62$ GeV. The solid line is the GM result and the dot-dashed
  is the results using the BGBK model (without Reggeon part). The
  long-dashed line is the GM results without Reggeon contribution.}
\label{fig5}
\end{figure}

As a final investigation, the $x_F$-integrated dilepton transverse
momentum distribution is calculated and compared with the available
data on $pp$ reactions at $\sqrt{s}=62 $ GeV and mass interval $5\leq
M\leq 8$ GeV (CERN R209) \cite{CERNR209}.  The results are presented in
Fig.~\ref{fig5}, with the solid curve denoting the Glauber-Mueller
calculation, including the non-asymptotic valence content (GM +
Reggeon), the dot-dashed line is the BGBK result \cite{newGBW} and the
long-dashed line is the Glauber-Mueller calculation without
non-asymptotic valence content (GM no Reggeon). The
calculation using the improved saturation model shows only fair agreement
with the experimental CERN data. Note however, that no reggeon part
has been introduced for the BGBK model. In addition, the data
shown in Fig.~\ref{fig5} were integrated over all $x_F$ and therefore
include contributions that are not taken into account by the dipole 
approach (see discussion in sect.~\ref{sec:relate}).
The GM result, on the other hand, 
is in good agreement with the overall normalization and
behavior presented by the data, when the non-asymptotic contribution is taken
into account, even though no parameters have been adjusted
to fit the data. 
The GM cross section 
overestimates the saturation model due to the inclusion of the
non-asymptotic contribution.

\section{Conclusions}

In this work, we investigated in detail the Drell-Yan transverse momentum
distribution in the color dipole framework, and we
analytically demonstrated that
the dipole approach correctly reproduces partially the NLO parton
model in the appropriate limit.
In contrast to the cross
section integrated over $p_T$, the DY $p_T$ distribution opens a kinematical
window where even large dipole configurations contribute. This
can be verified by studying the weight function associated with the
light cone wave functions for the process for 
different values of transverse momentum. Large partonic
configurations have their maximal contribution at $p_T=0$. A
remarkable feature of the dipole approach is the finite and well
behaved property of the dilepton $p_T$ distribution at $p_T\to 0$
in a LO calculation. 

The main motivation to pursue the dipole approach is that 
it provides a natural framework for the description of
unitarity effects, that are not taken into account by the
conventional parton model. Unitarity corrections are implemented 
in the dipole cross section, using the GM approach \cite{AGL}.
In addition, it is performed a comparison with the QCD improved
saturation model of the dipole cross section \cite{newGBW}.
In general, the unitarity corrections produce a
reduction in the differential cross section, mostly at large $p_T$. At
LHC energies, the corrections are quite large and they cannot be
reproduced by only using new adjustable parameterizations for the gluon
distribution.

In order to extrapolate the dipole approach to lower energies, a
Reggeon contribution was introduced  into the dipole cross section.
This Reggeon part is proportional to the valence quark
content of the target, meaning at high energies, i.e. RHIC and LHC,
it is negligible, although it is important in order to obtain 
a good description of the CERN ISR
data \cite{CERNR209}. 

\bigskip
{\bf Acknowledgments:}
This work was partially supported by CNPq (Brazil). J.R.\
was supported  by
the U.S.~Department of Energy at Los Alamos
National Laboratory under Contract No.~W-7405-ENG-38.

\end{document}